**Bulk magnetic order in a two dimensional $Ni^{1+}/Ni^{2+}$ ($d^9/d^8$) nickelate, isoelectronic with superconducting cuprates**

By *Viktor V. Poltavets, Konstantin A. Lokshin, Andriy H. Nevidomskyy, Mark Croft, Trevor A. Tyson, Joke Hadermann, Gustaaf Van Tendeloo, Takeshi Egami, Gabriel Kotliar, Nicholas ApRoberts-Warren, Adam P. Dioguardi, Nicholas J. Curro,* and *Martha Greenblatt\**

       Prof. Viktor V. Poltavets
Department of Chemistry and Chemical Biology, Rutgers University
Piscataway, NJ 08854 (USA) and Department of Chemistry, Michigan State University, East Lansing, MI 48824 (USA)
       Dr. Konstantin A. Lokshin, Prof. Takeshi Egami
Department of Materials Science and Engineering, University of Tennessee
Knoxville, TN 37996 (USA)
       Dr. Andriy H. Nevidomskyy, Prof. Mark Croft, Prof. Gabriel Kotliar
Department of Physics and Astronomy, Rutgers University
Piscataway, NJ 08854 (USA)
       Prof. Trevor A. Tyson
Department of Physics, New Jersey Institute of Technology
Newark, NJ 07102 (USA)
       Dr. Joke Hadermann, Prof. Gustaaf Van Tendeloo
EMAT, University of Antwerp
Antwerp B-2020 (Belgium)
       Nicholas ApRoberts-Warren, Adam P. Dioguardi, Prof. Nicholas J. Curro
Department of Physics, University of California
Davis, CA 95616 (USA)
[*]    Prof. Martha Greenblatt
Department of Chemistry and Chemical Biology, Rutgers University
Piscataway, NJ 08854 (USA)
E-mail: martha@rutchem.rutgers.edu

The $Ni^{1+}/Ni^{2+}$ states of nickelates have identical ($3d^9/3d^8$) electronic configuration as $Cu^{2+}/Cu^{3+}$ in the high temperature superconducting cuprates, and are expected to show interesting properties. An intriguing question is whether mimicking the electronic and structural features of cuprates would also result in superconductivity in nickelates? The $Ni^{1+}$ state is rare and cannot be easily stabilized. We have succeeded in synthesizing a family of such compounds, $Ln_{n+1}Ni_nO_{2n+2}$ with the T'-type structure, by low temperature reactions and have studied their properties. Here we report experimental evidence for a bulk-like magnetic transition in $La_4Ni_3O_8$ (n = 3) at 105 K. Density functional theory (DFT) calculations relate the transition to a spin density wave nesting instability of the Fermi surface.

Prior to their seminal discovery of high temperature superconductivity, Bednorz and Müller initially focused their search for superconductivity on the La-Ni-O system with Ni in 2+/3+ oxidation state.[1] After the discovery of superconductivity in the cuprates,[2] other transition metal oxides were investigated for superconductivity, and aside from the cobaltates no other system has been discovered to date. The nickelates, however, have the potential to exhibit physics similar to the cuprates. Theoretical studies show that only nickelates with $Ni^{1+}$ ($d^9$, S = 1/2) in a square planar coordination can form an antiferromagnetic (AFM) insulator directly analogous to the parent (undoped) cuprates.[3] Nickelates with $Ni^{1+}/Ni^{2+}$ are metastable and to date, there are only a few compounds known with infinite $Ni^{1+}/Ni^{2+}O_2$ planes with little known about their physical properties. Here we report experimental and theoretical studies which indicate that $La_4Ni_3O_8$, a two dimensional (2D) square planar nickelate with $Ni^{1+}/Ni^{2+}$, is antiferromagnetic below $T_N$ = 105. $La_4Ni_3O_8$ is a prime example of a layered compound, encompassing all the properties of the recently proposed hypothetical $LaNiO_3/LaMO_3$ (M = Al, Ga) superlattice oxide with alternating conducting $NiO_2$ and insulating $MO_2$ planes and theoretical predictions of favorable conditions for high $T_c$ superconductivity.[4]

$La_4Ni_3O_8$ is a member of the so-called $T'$-type $Ln_{n+1}Ni_nO_{2n+2}$ (Ln = La, Nd; n = 2, 3 and ∞) homologous series.[5, 6] The structures of the $T'$-type nickelates can be described by stacking of alternating (Ln/$O_2$/Ln) fluorite-type layers with $Ln_{n-1}(NiO_2)_n$ infinite layer structural blocks[5–7] (**Fig. 1**). Phases with the infinite layer structure (n = ∞) are known both for nickelates, $LnNiO_2$ (Ln = La, Nd),[8–13] and for cuprates, $Ca_{0.84}Sr_{0.16}CuO_2$ [ref. 14], whereas the so-called double (n = 2) and triple (n = 3) layer $T'$-type nickelates are unique examples of such structural arrangements.[5, 6]

The n = 3 member of the series, $La_4Ni_3O_8$, was first prepared by Laccore[15] in 1992 by low temperature reduction of Ruddlesden-Popper $La_4Ni_3O_{10}$ phase. The oxygens are removed exclusively from positions between NiO$_2$ layers in this process. In addition, a structural transformation from rock salt (LaO)$_2$ layers into fluorite La/O$_2$/La blocks occurs during the reaction. This so-called T-to-T' transformation is also known for cuprates.[16] It is noteworthy that even though La$_4$Ni$_3$O$_8$ is prepared by reduction, there are no oxygen vacancies; it is an oxygen stoichiometric phase. The distance between NiO$_2$ planes is 3.262(2) Å [ref. 13], which is too small to accommodate any oxygen atoms. Oxygen stoichiometry and the absence of local inhomogeneity make this phase an ideal model for experimental and theoretical investigations. Information on the physical properties of La$_4$Ni$_3$O$_8$ is limited to a remark that it is a black semiconductor, whereas magnetic and other physical properties have not been reported previously.[15] **Fig. 2a** shows the magnetization, $M$, of La$_4$Ni$_3$O$_8$ as a function of temperature. For large applied field ($H$ = 5 T), $M$ exhibits a sharp drop while cooling below $T$ = 105 K (Fig. 2a). This transition can be observed in magnetization measurements only at a high applied magnetic field ($H_{min}$ ~1000 Oe), while no sign of the transition is seen in $H$ = 100 Oe, even though thermodynamic measurements reveal the transition is also present in $H$ = 0, as discussed below. We attribute this anomalous field dependence to the presence of a small (less than 1%) admixture of superparamagnetic Ni particles, which appears to be present in all samples. The divergence between zero-field- and field-cooled curves at low temperature (< 50 K; Fig. 2a), the non-Curie-Weiss behavior in the high temperature region, which is deduced from the nonlinear shape of the $M$ vs. $H$ curves (Fig. S1), the nonlinear inverse magnetic susceptibility, $\chi^{-1}(T)$ (not shown) are also attributed to the small Ni impurity admixture. We found that while the value of magnetization anomaly at 105 K is reproducible for La$_4$Ni$_3$O$_8$ from different synthetic batches, the magnetization was strongly sample dependent. However, only La$_4$Ni$_3$O$_8$ was found in all samples by powder X-ray diffraction, as well as by neutron powder diffraction (NPD). High resolution transmission electron microscopy showed that the crystallites themselves are also single phase down to nanometer scale (Fig. S2).

The temperature dependence of the resistivity (Fig. 2b) reveals semiconducting behavior. Moreover, a clear change of slope can be seen at 105 K in Fig. 2b. The transition temperature (~ 105 K) is consistent with the results of magnetic measurements, and supports the intrinsic character of the transition in $La_4Ni_3O_8$. The resistivity values measured at $H = 5$ T coincide, within experimental error, with those measured in zero field. Therefore we conclude that the transition seen at 105 K is also present at $H = 0$, but it is only detectable for $H$ above the saturation magnetization of the admixed Ni particles, i.e. above 1000 Oe. The resistivity data below the 105 K regions is clearly insulating and can be fit to Mott's law for variable range hopping (Fig. 2b, inset).[17] Above the transition, the resistivity is consistent with either a bad metal, or an Anderson insulator. It should be noted that due to the metastability of $La_4Ni_3O_8$ (the sample decomposes at T > 350ºC), the transport measurement could only be performed on a polycrystalline pressed pellet, where poor intergrain electrical contacts are expected. Heat capacity data for $La_4Ni_3O_8$ measured at $H = 0$ T are presented in Figure 2c by filled-red circles. The lambda anomaly, peaking at 105 K, is further evidence for a bulk phase transition at this temperature. In order to estimate the entropy loss during the transition, the scaled data for the specific heat capacity of the closely related phase, $La_3Ni_2O_6$ were subtracted from those of $La_4Ni_3O_8$, and the area under the resultant peak was integrated (Fig. 2c). The experimental value of $\Delta S_{exp}$ (per 1 mol of $La_4Ni_3O_8$) = 5.96 J/(mol K). The large value of the entropy loss asserts the transition to be intrinsic to $La_4Ni_3O_8$. $\Delta S_{exp}$ is close to Rln2 = 5.76 J/(mol K). In a localized model this value is close to 1/2 Rln2 per $Ni^{1+}$, or 1/3 Rln2 per Ni in an itinerant picture.

Thus magnetic, resistivity and heat capacity measurements all indicate a phase transition at 105 K. The NPD pattern of $La_4Ni_3O_8$ at 15 K is identical to that at room temperature, except for a thermal contraction and expected difference in thermal factors, which indicate that this transition cannot involve a large structural displacement. Extended X-ray absorption fine structure (EXAFS) measurements support this conclusion. In addition the EXAFS data show a sharp drop of the mean squared relative displacement ($\sigma^2$) values for the average Ni-O bond distance near ~105 K, very near the critical point, followed by a recovery at higher temperatures to the low-temperature value (Fig. S3). It is interesting to note that recent EXAFS local structure measurements[18] showed the same behavior of ($\sigma^2$) for in-plane bonds at the spin density wave (SDW) ordering temperature in $Ca_3Co_4O_9$, a 2D magnetic oxide.[19] In order to investigate the phase transition in more detail, we have carried out $^{139}$La nuclear magnetic resonance (NMR; I = 7/2) in an aligned powder sample to probe the spin dynamics and internal hyperfine fields associated with the low temperature state. We identify two sites with electric field gradients (EFGs) of 1.35(5) MHz and ~20(10) kHz, and assign these to La(1) (between $NiO_2$ planes - see Fig. 1) and La(2) (in the fluorite blocks), respectively based on EFG calculations with Wien2K.[20] Below the phase transition, the La(1) spectrum is washed out by the presence of a broad distribution of static hyperfine fields, whereas the La(2) resonance remains visible (see Fig. S7). The powder nature of the sample coupled with the small value of the La(2) EFG (~20kHz) renders the La(2) spectrum nearly featureless, but we can identify two sub-peaks in the spectrum which develop a significant temperature dependence below 105 K (see Fig. S8). We find an internal field ~200 Oe develops at the La(2) site with a second-order mean-field like temperature dependence. This internal field arises from the non-cancellation of the transferred hyperfine fields at the La(2) site from ordered Ni moments, and is thus a direct measure of the order parameter of this system. A similar field develops at the La(1) site, but since the hyperfine fields are much larger in the La(1) case, the spectrum is broadened considerably. In **Fig. 3** we show the nuclear spin lattice relaxation rate at the two sites, which clearly shows a dramatic suppression below 105 K and confirms our interpretation that this is an intrinsic phenomenon. At high temperatures, $(T_1T)^{-1}$ approaches a constant for both sites, indicative of Fermi liquid behavior. Between 250 K and 105K $(T_1T)^{-1}$ increases dramatically, revealing critical spin-fluctuations associated with AFM transition.[21] Both La(1) and La(2) exhibit similar temperature dependences for $(T_1T)^{-1}$, although $(T_1T)^{-1}$ is larger for the La(1) because of the larger hyperfine coupling to the Ni moments at this site. The behavior of the La(1) and La(2) are thus analogous to the Cu and Y nuclei in $YBa_2Cu_3O_7$ [ref. 22]. Below the ordering temperature $(T_1T)^{-1}$ once again becomes independent of temperature with a reduced value. This observation confirms our DFT calculations that only a portion of the density of states at the Fermi level is gapped out and is analogous to the observed behavior in other systems, such as $URu_2Si_2$ [ref. 23].

In order to gain insight on the nature of the phase transition at 105 K we have determined the electronic band structure and Fermi surface (FS) with the full potential linearized augmented plane wave method as implemented in the Wien2k package[20] (**Fig. 4**). There are three FS's with strong 2D character. The sheet centered at the $\Gamma$ point has a square shape that could support a charge (CDW) or SDW instability close to the nesting wave vector of $Q = [1/3, 1/3, L]$. FS nesting resulting in CDW or SDW transitions was shown earlier for many low-dimensional transitional metal oxides.[24] In particular, incommensurate SDW instabilities are universal for $La_2CuO_4$-based superconductors.[25] For a SDW transition, only the magnetic translational symmetry is broken and no structural transition is required. On the other hand a CDW onset is always accompanied by a structural transition due to the coupling of the CDW with the crystal lattice. The absence of any change in the NPD pattern through the transition (i.e., no extra peaks, and no change in symmetry) excludes the possibility that the transition at 105 K is due to formation of a CDW state (Fig. S4 and Table S1). However, a SDW state usually results in extra magnetic peaks that should be observable by neutron scattering. The absence of magnetic peaks in the low temperature NPD pattern suggests that the system may not be fully three-dimensionally ordered. The FS shown in Fig. 4 indicates a very strong 2D character of the electronic structure. The structure of this compound is composed of three $NiO_2$ layers separated by a $La/O_2/La$ layers (Fig. 1). Even though the magnetic interactions within the tri-layers could be significant, the interaction between different tri-layers must be very weak. In addition the Ni position is shifted by [1/2, 1/2] across the $La/O_2/La$ layer, frustrating the antiferromagnetic order. Consequently there is a strong possibility that the ordering at 105 K occurs only within the tri-layer. The magnetic ordering peaks will then be broad along $L$ ($c^*$-axis), making the detection by powder diffraction extremely difficult.

Additional DFT (LDA and LDA+$U$[26]) calculations have been carried out to determine the spin configuration of the ground state and the electronic structure of the material (calculation details are available in the Supporting Information). We find that the proposed SDW state with ordering wave-vector $Q$ = [1/3, 1/3, 0] has a lower ground state energy than the paramagnetic state, confirming our previous conclusions (see Supporting Information for details). The Hubbard on-site interaction for Ni d-electrons, $U$ = 4.3 eV, was determined by constrained DFT calculations; the $U$ value characterizes $La_4Ni_3O_8$ as a material with intermediate electron correlations. The value of $U$ in this nickelate is considerably smaller than in the cuprates (9-11 eV)[27] and notably larger than in the iron-pnictide superconductor $LaFeAsO_{1-x}F_x$ (1.5 eV).[28] According to our DFT calculations with onsite correlations taken into account within the local-density approximations (LDA)+$U$ method, $La_4Ni_3O_8$ remains metallic. Also due to the highly anisotropic crystal structure of $La_4Ni_3O_8$ (Fig. 1) and the Fermi surface (**Fig. 4**), large differences in $ab$-plane and in $c$-direction conductivity can be expected: i.e., metallic conductivity in the infinite $Ni^{1+/2+}O_2$ planes, while semiconducting or insulating behavior perpendicular to those planes. Evidently, the semiconducting property is dominant (Fig. 2b) due to poor percolation in the $NiO_2$ planes of the poorly compacted polycrystalline sample.

To summarize, a magnetic transition at 105 K was established for $La_4Ni_3O_8$ by magnetization, resistivity, specific heat, NPD, EXAFS and $^{139}La$ NMR methods. Absence of a structural transition and minute $c/a$ ratio changes between 15 K and 300 K argue against CDW or $Ni^{2+}$ ($d^8$) high-spin to low spin transitions. Thus the spin ordering or fluctuations are apparently of an AFM type. This is supported by the DFT calculations, which relate the transition to a SDW nesting instability of the FS with ordering wave-vector $Q$ = [1/3, 1/3, L] with $L \approx 0$. We note that the fact that $(T_1T)^{-1}$ begins to increase at temperatures on the order of $3T_N$ strongly suggests that the spin fluctuations are 2D, confined to the $NiO_2$ trilayer, and may indicate the presence of a smaller interlayer coupling as the origin of the long-range magnetic order. Furthermore, this strong temperature dependence of $(T_1T)^{-1}$ is similar to that observed in several unconventional superconductors, and suggests that $La_4Ni_3O_8$ may become superconducting upon appropriate doping. Once again synthetic solid state chemistry lead to the realization of a bulk mixed valent Ni-layered compound, similar to the theoretically predicted high $T_c$ superconducting artificial $LaNiO_3/LaAlO_3$ superlattice phase.[4]

*Experimental*

The preparation of $La_4Ni_3O_{10}$ and $La_4Ni_3O_8$ is described in [13]. Details of NPD, magnetic, resistivity, specific heat measurement and Rietveld refinement are available in the Supporting Information. The EXAFS data were collected at beamline X-11A at the national synchrotron light source (NSLS), Brookhaven National Laboratory. Details of the measurements can be found in [18]. The electronic structure calculations were performed by the full potential linearized augmented plane wave method as implemented in the Wien2k package [20] with on-site correlations taken into account [26]. Further calculation details are available in the Supporting Information.

*Acknowledgements*
This work was supported by the National Science Foundation through NSF-DMR-0541911 (V.V.P., M.G) and NSF-DMR-0528969 (G.K.). T.A.T. acknowledges support from U. S. Department of Energy Office of Basic Energy Sciences (DOE-BES) Grant DE-FG02-07ER46402. K.A.L. and T.E. acknowledge support from DOE-BES EPSCoR Implementation Award DE-FG02-08ER46528. Use of the National Synchrotron Light Source, Brookhaven National Laboratory, was supported by DOE-BES under Contract No. DE-AC02-98CH10886. Supporting Information is available online from Wiley InterScience or from the authors.

**Figure 1.** Crystal structures of $La_4Ni_3O_{10}$ and $La_4Ni_3O_8$ with denoted layers and structural blocks: P - perovskite, RS - rock salt, IL - infinite layer, F - fluorite.

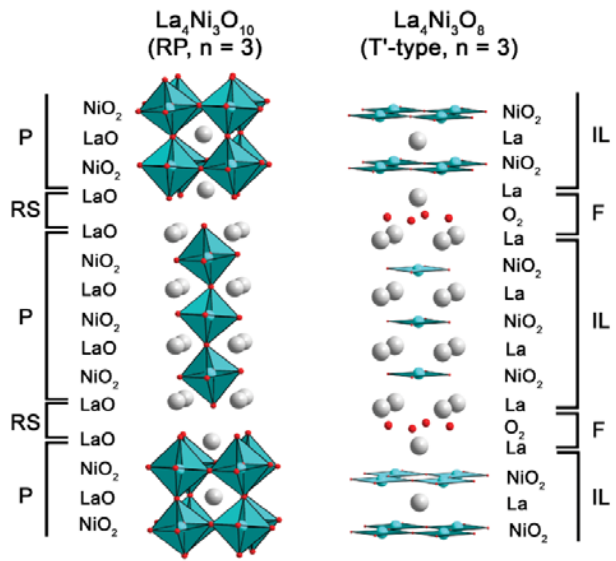

**Figure 2.** Temperature dependence of selected physical properties: a) Mass magnetization $M$ vs. temperature ($T$) for $La_4Ni_3O_8$ measured in different magnetic fields (ZFC - zero field cooling, FC - field cooling). b) Temperature dependence of resistivity (logarithmic scale). The inset shows the natural logarithm of resistivity, $\ln(\rho)$ versus $T^{(-1/4)}$: observed data (red circles) and line fits (blue solid lines) according to Mott's variable-range-hopping model. c) Specific heat data of $La_4Ni_3O_8$ (red dots) and $La_3Ni_2O_6$ (green dots). The dashed blue line represents the scaled specific heat of $La_3Ni_2O_6$.

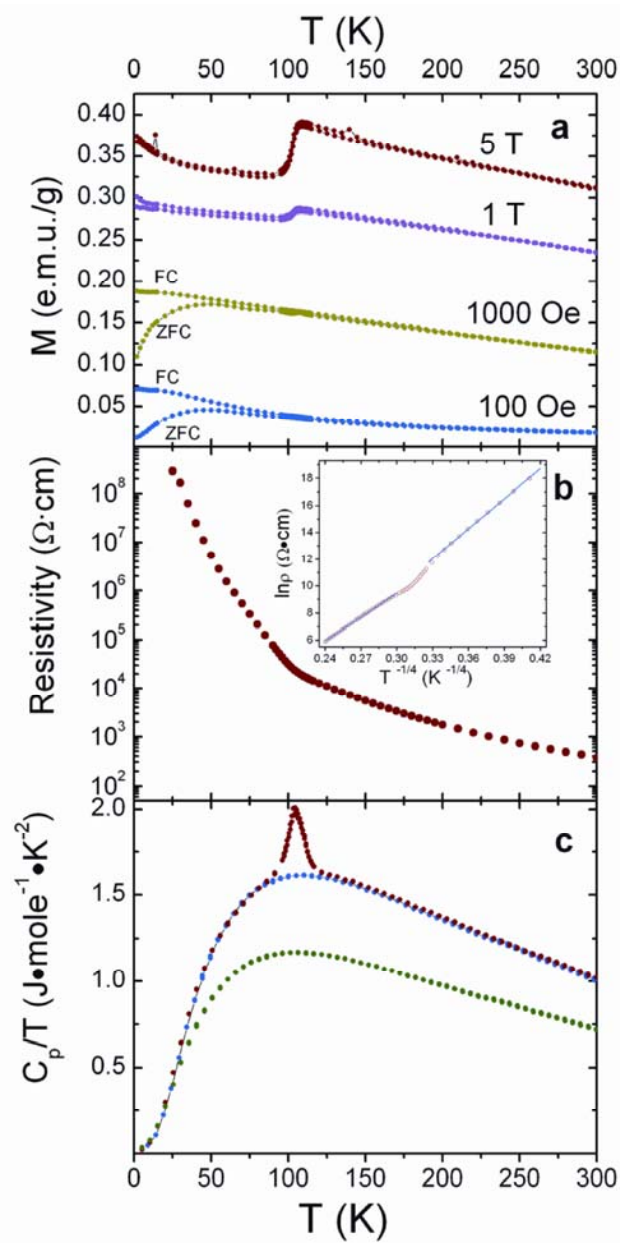

**Figure 3.** The nuclear spin lattice relaxation rate of the $^{139}$La versus temperature, measured at the 3/2 -5/2 satellite of the La(1) (blue circles) and at the main line (La(1) and La(2), red circles). INSET: $^{139}$La NMR spectrum in an aligned powder sample at 200 K at fixed frequency of 25.1 MHz, revealing both the axially symmetric La(1) and La(2). The arrows indicate the fields where the T$_1$ data was obtained.

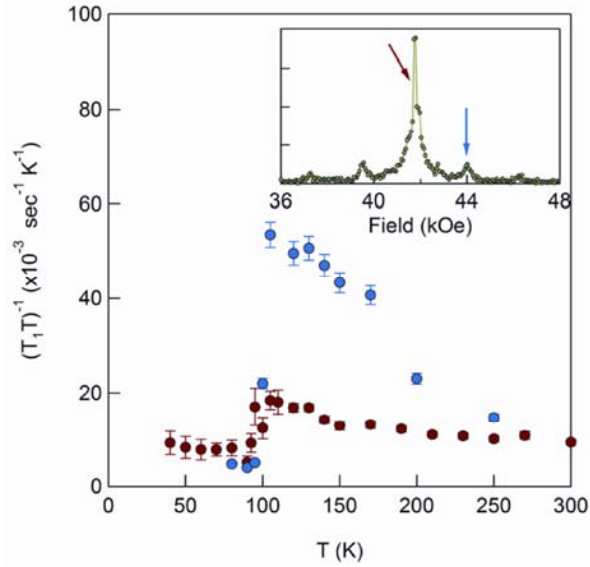

**Figure 4.** Fermi surface of La$_4$Ni$_3$O$_8$ in reciprocal space with nesting vectors (pink arrows).

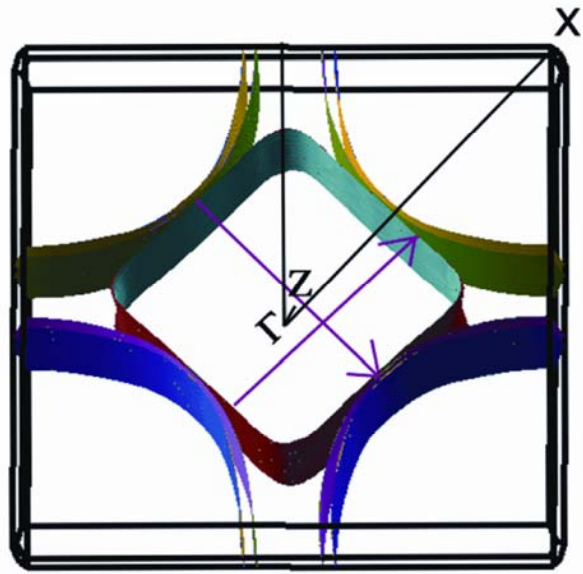

**The table of contents entry**.

Bulk magnetic order in two dimensional $La_4Ni_3O_8$ nickelate with $Ni^{1+}/Ni^{2+}$ ($d^9/d^8$), isoelectronic with superconducting cuprates is demonstrated experimentally and theoretically. Magnetization, specific heat and $^{139}La$ NMR evidence a transition at 105 K to an antiferromagnetic state. Theoretical calculations by DFT relate the transition to a nesting instability of the Fermi surface with ordering wave-vector Q = [1/3, 1/3, 0].

Viktor V. Poltavets, Konstantin A. Lokshin, Andriy H. Nevidomskyy, Mark Croft, Trevor A. Tyson, Joke Hadermann, Gustaaf Van Tendeloo, Takeshi Egami, Gabriel Kotliar, Nicholas ApRoberts-Warren, Adam P. Dioguardi, Nicholas J. Curro, Martha Greenblatt*

**Bulk magnetic order in a two dimensional $Ni^{1+}/Ni^{2+}$ ($d^9/d^8$) nickelate, isoelectronic with superconducting cuprates**

ToC figure

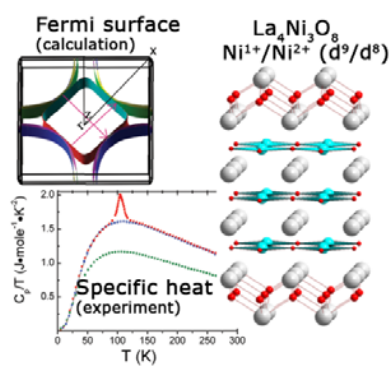

# Supporting Information

**Bulk magnetic order in a two dimensional $Ni^{1+}/Ni^{2+}$ ($d^9/d^8$) nickelate, isoelectronic with superconducting cuprates**

By *Viktor V. Poltavets, Konstantin A. Lokshin, Andriy H. Nevidomskyy, Mark Croft, Trevor A. Tyson, Joke Hadermann, Gustaaf Van Tendeloo, Takeshi Egami, Gabriel Kotliar, Nicholas ApRoberts-Warren, Adam P. Dioguardi, Nicholas J. Curro*, and *Martha Greenblatt\**

Prof. V. V. Poltavets
Department of Chemistry and Chemical Biology, Rutgers University
Piscataway, NJ 08854 (USA) and Department of Chemistry, Michigan State University, East Lansing, MI 48824 (USA)

Dr. K. A. Lokshin, Prof. Takeshi Egami
Department of Materials Science and Engineering, University of Tennessee
Knoxville, TN 37996 (USA)

Dr. A. H. Nevidomskyy, Prof. M. Croft, Prof. Gabriel Kotliar
Department of Physics and Astronomy, Rutgers University
Piscataway, NJ 08854 (USA)

Prof. T. A. Tyson
Department of Physics, New Jersey Institute of Technology
Newark, NJ 07102 (USA)

Dr. J. Hadermann, Prof. G. V. Tendeloo
EMAT, University of Antwerp
Antwerp B-2020 (Belgium)

N. ApRoberts-Warren, A. P. Dioguardi, Prof. N. J. Curro
Department of Physics, University of California
Davis, CA 95616 (USA)

[*]   Prof. Martha Greenblatt
Department of Chemistry and Chemical Biology, Rutgers University
Piscataway, NJ 08854 (USA)
E-mail: martha@rutchem.rutgers.edu

## Experimental

### General

The dc magnetic susceptibility measurements were carried out on powder samples with a commercial SQUID magnetometer (Quantum Design, MPMS-XL). All measurements were performed by warming the samples in the applied field after cooling to 5 K in zero field (zero field cooling) and by cooling the samples in the applied measuring field (field cooling). Resistivity measurements were performed on the same machine with Keithley equipment and with a standard four-probe technique. The EXAFS data were collected at beamline X-11A at the national synchrotron light source (NSLS), Brookhaven National Laboratory. The specific heat was measured in the temperature range 5–300 K at zero magnetic field by relaxation calorimetry. This method has a relative accuracy of ±3%.

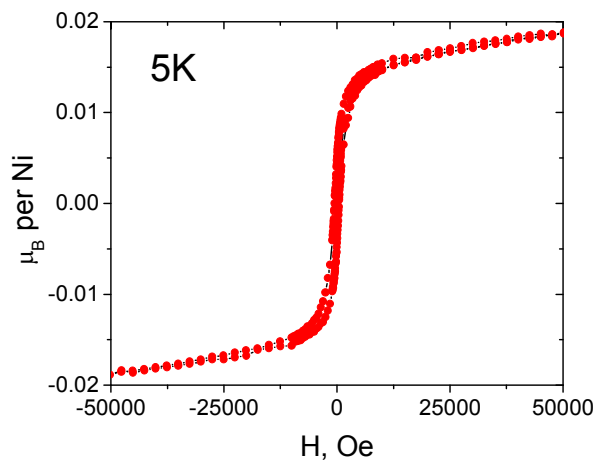

**Fig. S1.** Magnetic moment $\mu_B$ (Bohr magneton) per Ni in $La_4Ni_3O_8$. The presence of the small amount of Ni admixture is ignored and all the magnetic response is ascribed to $La_4Ni_3O_8$.

### Electron Microscopy

The sample for transmission electron microscopy (TEM) investigation was prepared by crushing the powder sample in ethanol and depositing it on a holey carbon grid. For high resolution transmission electron microscopy a JEOL 4000EX microscope was used. Scanning electron microscopy (SEM) images were taken from the uncrushed powder, using Nova 200 Nanolab.

According to the scanning electron microcopy data (Fig. S2a), the average crystallite size of $La_4Ni_3O_8$ lies in the range of 1 to 5 micrometers. High resolution transmission electron microscopy showed all studied crystallites to be homogenous without any formation of domains down to nanoscale (Fig. S2b).

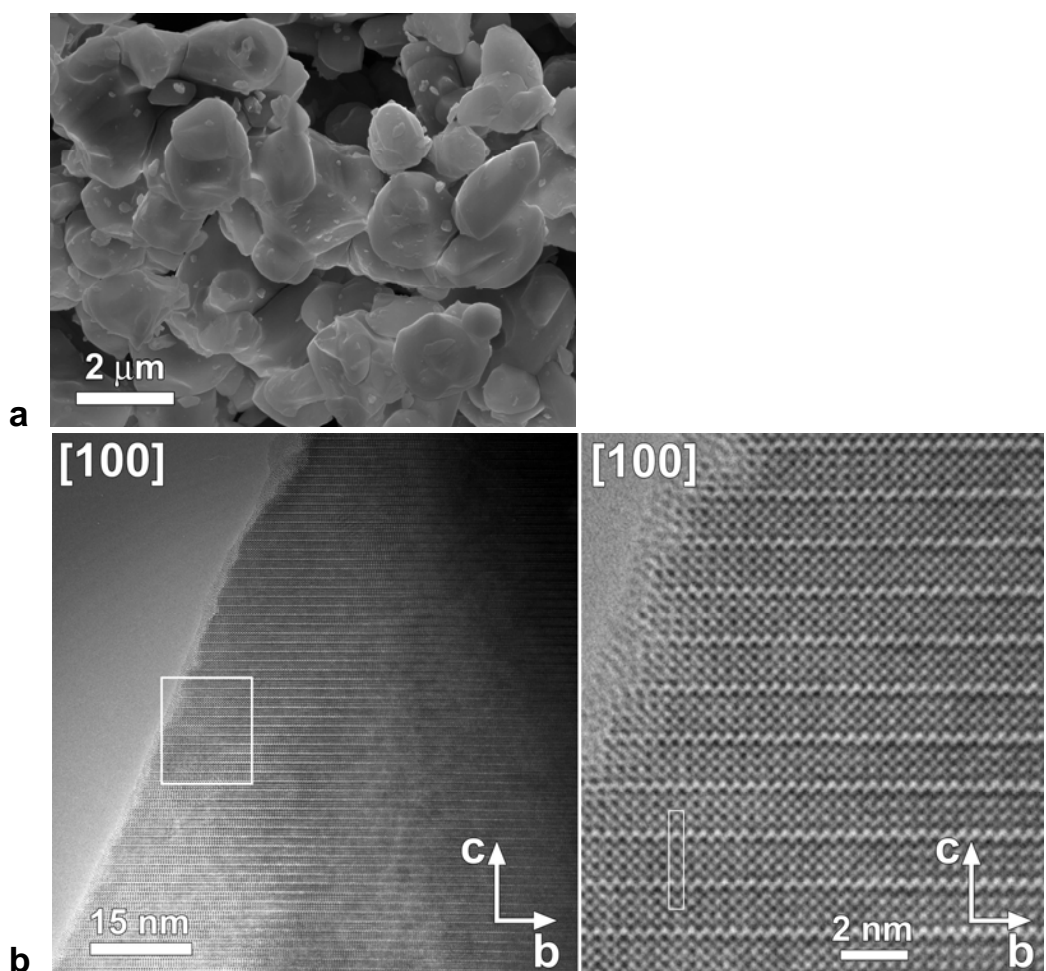

**Fig. S2.** Electron microscopy images. a) Scanning electron microscopy image of $La_4Ni_3O_8$. b) High resolution transmission electron microscopy image of $La_4Ni_3O_8$, left side indicates an overview of an area of approximately 150 nm by 100 nm, which shows neither defects nor domains. The area indicated by a white rectangle in the overview image is shown enlarged at the right side.

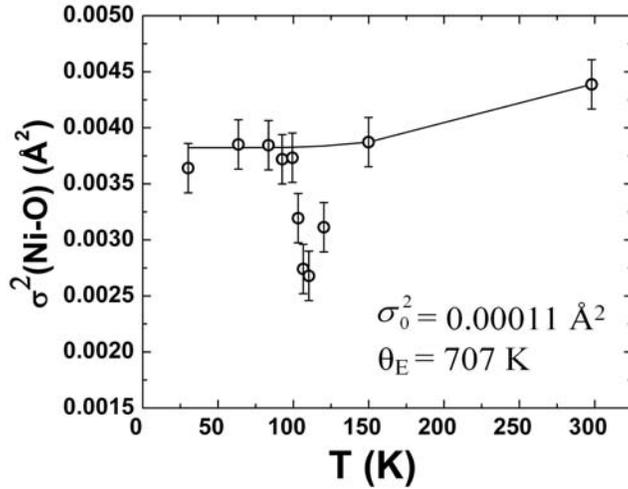

**Fig. S3.** Gaussian width ($\sigma^2 = \langle(R - \langle R\rangle)^2\rangle$ giving the mean squared relative displacement of a bond)) for the average Ni-O bond shown as circles. The extracted widths at 300 K and 150 K were fit to the Einstein model $\sigma^2(T)=\sigma_0^2+\dfrac{\hbar^2}{2\mu k_B \theta_E}\coth(\dfrac{\theta_E}{2T})$ and extrapolated to lower temperatures (line). The dip in $\sigma^2$ (T) reveals enhanced Ni-O correlations at the SDW ordering temperature.

## Neutron Powder Diffraction Experiment

Neutron powder diffraction data were collected on a 1g sample at 300 K and on a 2 g sample from another synthetic batch at 15 K on the NPDF time-of-flight neutron diffractometer at the Lujan Neutron Science Center of the Los Alamos National Laboratory. Rietveld refinements [S1] of the obtained data were performed with GSAS [S2] program with EXPGUI [S3] interface. In the final runs, the scale factor, unit-cell parameters, absorption coefficient, atomic coordinates and isotropic atomic displacement parameters were simultaneously refined. The preparation of $La_4Ni_3O_{10}$ and $La_4Ni_3O_8$ as well as the room temperature NPD data for $La_4Ni_3O_8$ were published earlier [S4].

**Table S1.** Crystallographic data for $La_4Ni_3O_8$ obtained by the Rietveld refinement of the neutron powder diffraction data measured at 15 K.

| Atom | Wyckoff position | x | y | z | $10^2 U_{iso}$ (Å) | Occupancy |
|------|------------------|---|---|---|--------------------|-----------|
| La1  | 4e | 0 | 0   | 0.4333(1) | 0.44(1) | 1 |
| La2  | 4e | 0 | 0   | 0.2996(1) | 0.39(1) | 1 |
| Ni1  | 2a | 0 | 0   | 0         | 0.52(1) | 1 |
| Ni2  | 4e | 0 | 0   | 0.1248(1) | 0.51(1) | 1 |
| O1   | 4c | 0 | 0.5 | 0         | 0.61(1) | 1 |
| O2   | 8g | 0 | 0.5 | 0.1261(1) | 0.72(1) | 1 |
| O3   | 4d | 0 | 0.5 | 0.25      | 0.44(1) | 1 |

Note. Space group *I4/mmm* (No. 139). Lattice parameters (Å): $a = b = 3.9633(1)$, $c = 26.0373(3)$. $\chi^2 =$ 3.0 %, wRp = 6.5 %, Rp = 4.8 %.

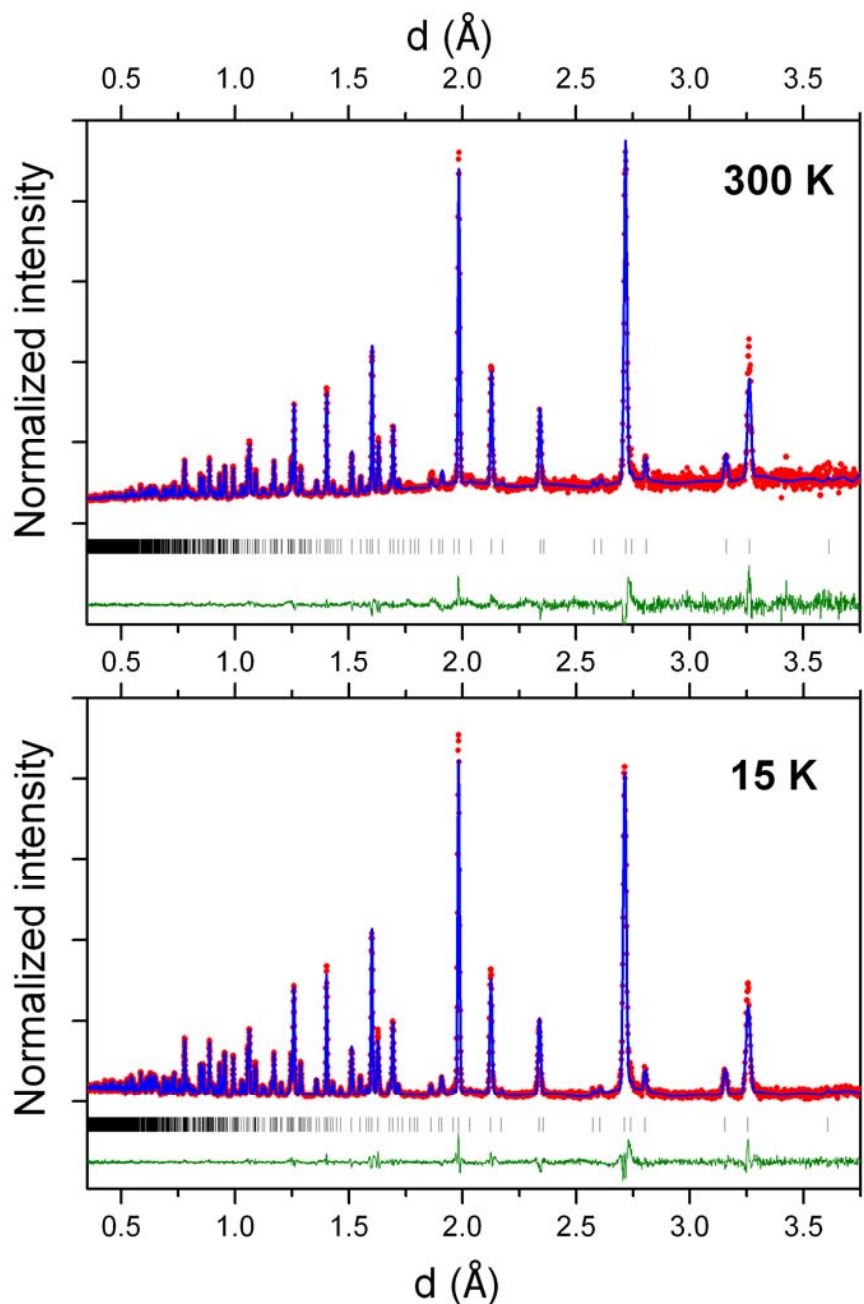

**Fig. S4.** Rietveld refinement profiles: observed (red dots), calculated (blue solid line), difference (green solid line) and Bragg reflections (tick marks) for the fit to the NPD patterns of $La_4Ni_3O_8$.

**Electronic structure calculations**

The electronic structure was calculated by the full potential linearized augmented plane wave (FLAPW) method as implemented in Wien2k [S5]. The exchange-correlations functional was taken within the generalized gradient approximation (GGA) in the parameterization of Perdew, Burke, and Ernzerhof (PBE) [S6]. To account for the most important on-site correlations the PBE-GGA + Hubbard $U$ approach in a Hartree-Fock-like scheme was used [S7, S8]. For the atomic spheres, the muffin-tin radii ($R_{MT}$) were chosen as 2.35 a.u, 1.97 a.u. and 1.75 a.u for the La, Ni and O sites, respectively. The number of plane waves was limited by a cut off $R_{MT}K_{max} = 7$.

There are 2 nonequivalent Ni positions in the $La_4Ni_3O_8$ crystal structure. The Ni1 site is in the $NiO_2$ planes between La layers. The Ni2 site is in the $NiO_2$ planes adjacent to $La/O_2/La$ fluorite blocks of the structure (Fig. 1 in the article). There are twice as many Ni2 as Ni1 positions per formula unit. The ferromagnetic (FM) results were derived from an initially parallel orientation of the spins at all the Ni atoms. An ordered state with ferromagnetic spin arrangement in the $NiO_2$ planes and antiferromagnetic coupling between layers will be denoted as AFM-BL (antiferromagnetic between layers) in further consideration. To allow antiferromagnetic checkerboard (AFM-CB) configuration, the crystal structure was transformed to *Ammm* space group (SG). The choice of the spin density wave (AFM-SDW) structure in *I4/mmm* SG with tripled *a* cell parameter was based on the FS nesting vector **Q** = [1/3, 1/3, 0] (see the article text). The transformation results in the splitting of each symmetrically independent Ni position into 3 new ones as it shown in Fig. S5. The initial polarization of the La and O atoms was set to zero, when starting the self-consistent cycles. The converged spin polarized calculations resulted in very small, induced spin moments at the O atoms that will be neglected in the discussion.

The Brillouin zone was sampled on a tetrahedral mesh with: 1000 k points (102 k points in the irreducible wedge of the Brillouin zone (IBZ)) for Ferro and AFM-BL; 155 k points in IBZ for AFM-CB calculations. For the AFM-SDW structure with large $3a \times 3a \times c$ supercells, a $3 \times 3 \times 3$ point mesh was used as base for the integration, resulting in 6 k-points in the IBZ. The energy convergence criterion was set in all calculations to $10^{-5}$ Ry (0.136 meV) and simultaneously the criterion for charge convergence was set to $10^{-3}$ e.

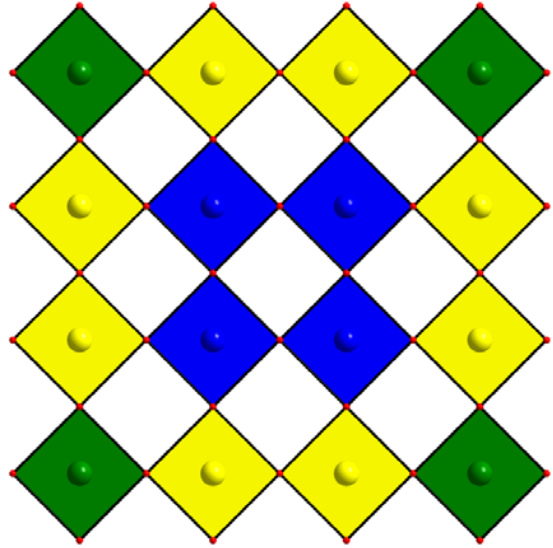

**Fig. S5.** Coordination polyhedra for symmetrically independent Ni positions in one Ni(1)O$_2$ layer of AFM-SDW configuration: Ni1-1 –green, Ni1-2 –yellow, Ni1-3 –blue.

In all the performed calculations, the total charges in the muffin-tin spheres had diminutive differences for crystallographically independent Ni positions, thus excluding a CDW scenario in agreement with the experimental data. It was found by GGA-PBE calculations (without $U$ applied) that the total energy of the AFM-SDW state was 9.23 meV (per Ni) lower than that of the paramagnetic state. The AFM-BL total energy is 14 meV lower than the AFM-SDW state. To assess effects of intra-atomic repulsion LDA+$U$ "correlated band theory" method was applied. The effective Hubbard $U$, $U_{eff}$ = 4.3 eV was derived on the basis of *ab initio* constrained calculations [S9]. This value was used in all calculations with correlations included. In LDA+$U$ calculations the AFM-BL and FM configurations are virtually degenerate and have the lowest total energy; however, they can be safely ruled out as they both would have a large net magnetization per unit cell, which was not observed experimentally. Only a very small magnetic moment per Ni can be deduced from the *M* vs. *H* measurements, even if the presence of the small amount of Ni admixture is ignored and all the magnetic response is ascribed to $La_4Ni_3O_8$ (Fig. S1). The AFM Neel state, AFM-CB, does not suffer from these drawbacks, but the calculated Fermi surface of $La_4Ni_3O_8$ does not have a nesting instability at the Neel wave-vector, Q = [1/2,1/2,0], and moreover the ground state energy of such a configuration was found to be substantially higher than that of the proposed AFM-SDW state. Therefore, even though it is not the lowest calculated total energy, the AFM-SDW is the only viable spin configuration. Obviously both LDA and LDA+*U* overestimated the tendency toward FM and AFM-BL states. It was proposed for LDA that this tendency can be corrected by accounting for magnetic fluctuations [S10]. It is also known, that even in the case of Cr, a classical SDW material, neither LDA nor GGA calculations predict a stable SDW ground state [S11]. The LDA+*U* approach is known to work well only in the case of integer orbital occupation and well-established long-range magnetic order [S12], which is not the case for partially filled Ni d-bands in $La_4Ni_3O_8$. An important feature of the AFM-SDW DOS is the appearance of a partial gap at the Fermi energy due to the removal of part of the Fermi surface below the transition (Fig. S6). Such gap is absent in other considered configurations.

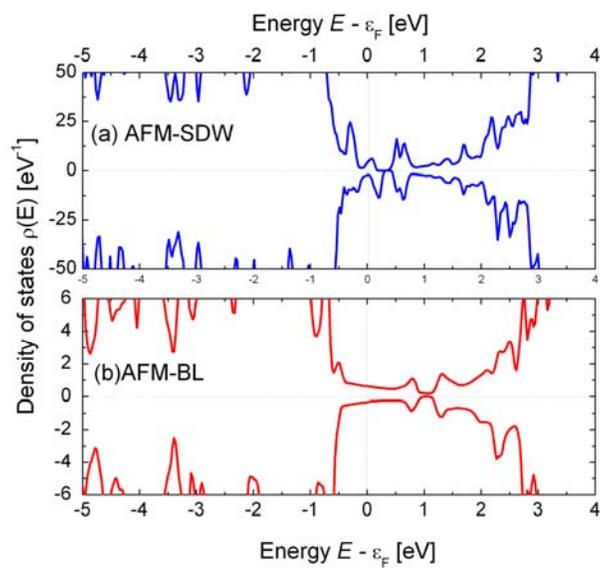

**Fig. S6.** LDA+$U$ total density of states for a) AFM-SDW and b) AFM-BL configurations.

**Table S2.** Results of LDA +$U$ ($U$ = 4.3 eV) calculations for magnetically ordered states of La$_4$Ni$_3$O$_8$.

| Spin configurations | ΔE per Ni relative to AFM-BL state, meV | Total magnetic moment per La$_4$Ni$_3$O$_8$, μ$_B$ | Ni spin magnetic moment, μ$_B$ | |
|---|---|---|---|---|
| | | | Ni1 | Ni2 |
| AFM-BL (FM orientation in NiO$_2$ layers, AFM between layers) | 0 | 0.64 | -0.80 | 0.80 |
| Ferromagnetic | 2.34 | 2.00 | 0.77 | 0.75 |
| AFM-SDW | 63.48 | 0.04 | 0.75; 0.73; -0.83 | -0.74; -0.66; 0.82 |
| AFM-CB (AFM checkerboard in NiO$_2$ layers) | 211.44 | 0 | -0.96; 0.96 | 0.63; -0.63 |

**Nuclear Magnetic Resonance**

We performed $^{139}$La (I=7/2) on both random and aligned powders of of La$_4$Ni$_3$O$_8$. The aligned powder was prepared by mixing powder sample with Stycast 1266 epoxy and curing in an external field of 9 T. Field-swept spectra were obtained by integrating the spin-echo intensity as a function of applied field at fixed frequency. By comparing the powder pattern of the random powder and the spectra of the aligned, we found two La sites: one with a quadrupolar splitting of $\nu_Q$ = 1.35(5) MHz with the principal axis of the electric field gradient (EFG) tensor along the alignment axis and the other with $\nu_Q$ ~ 20(10) kHz. A priori we cannot determine if the alignment axis is along the c or ab direction, but by symmetry we expect the EFG axis (and thus the alignment direction) to lie along c. By comparing the calculated EFG's determined by Wien 2K GGA methods with our observed spectra, we identify the La(1) site (between NiO$_2$ planes) with the larger splitting and the La(2) (in the fluorite blocks) with the smaller. The spin lattice relaxation rate was measured by inversion recovery, and the magnetization was fit to the standard expression for the particular transition for a spin 7/2 nucleus in order to extract a value for $T_1$. Field-swept spectra are shown in Fig. S7. Below the ordering temperature, we clearly see changes in both the La(1) and La(2) sites. Whereas the La(1) is wiped out due to the broad distribution of static internal hyperfine fields, the La(2) resonance remains visible. Although the quadrupolar satellite structure of the La(2) is poorly resolved, there are sub-peaks that shift visibly below 100K. We attribute this shift of the peaks to the development of small static internal hyperfine fields from the magnetic order (on the order of +/- 200 Gauss). The temperature dependence of these sub-peaks is shown in Fig. S8, which clearly shows changes indicative of a second order mean-field-like transition.

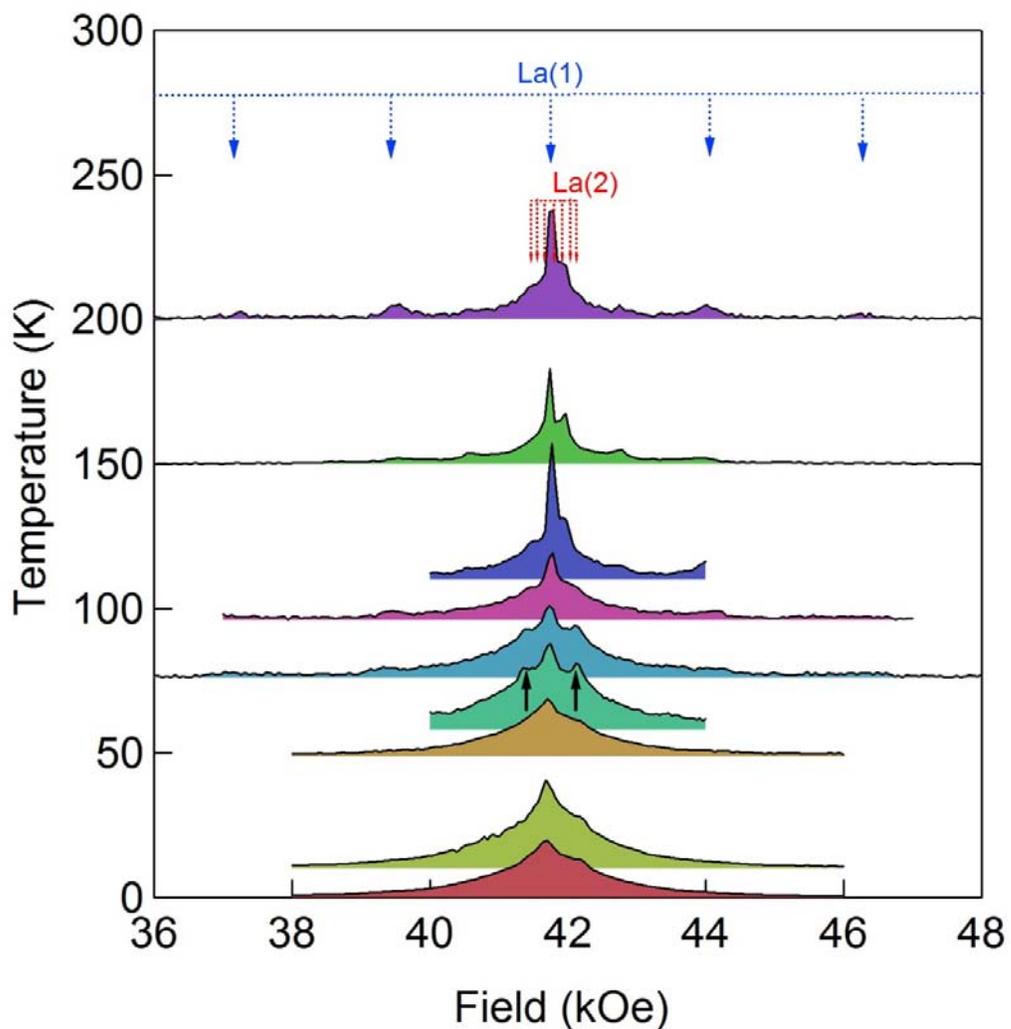

**Fig. S7**. Field-swept NMR spectra as a function of temperature at a fixed frequency of 25.1 MHz in an aligned powder of $La_4Ni_3O_8$. The La(1) and La(2) resonances are labelled, and the position of the temperature dependent peaks plotted in Fig. S8 are shown as black arrows for the 60 K spectra. At low temperatures, the La(1) sites are so broadened by the distribution of internal fields that their overall intensity is wiped out. The La(2) sites are broadened, but remain visible down to 4K.

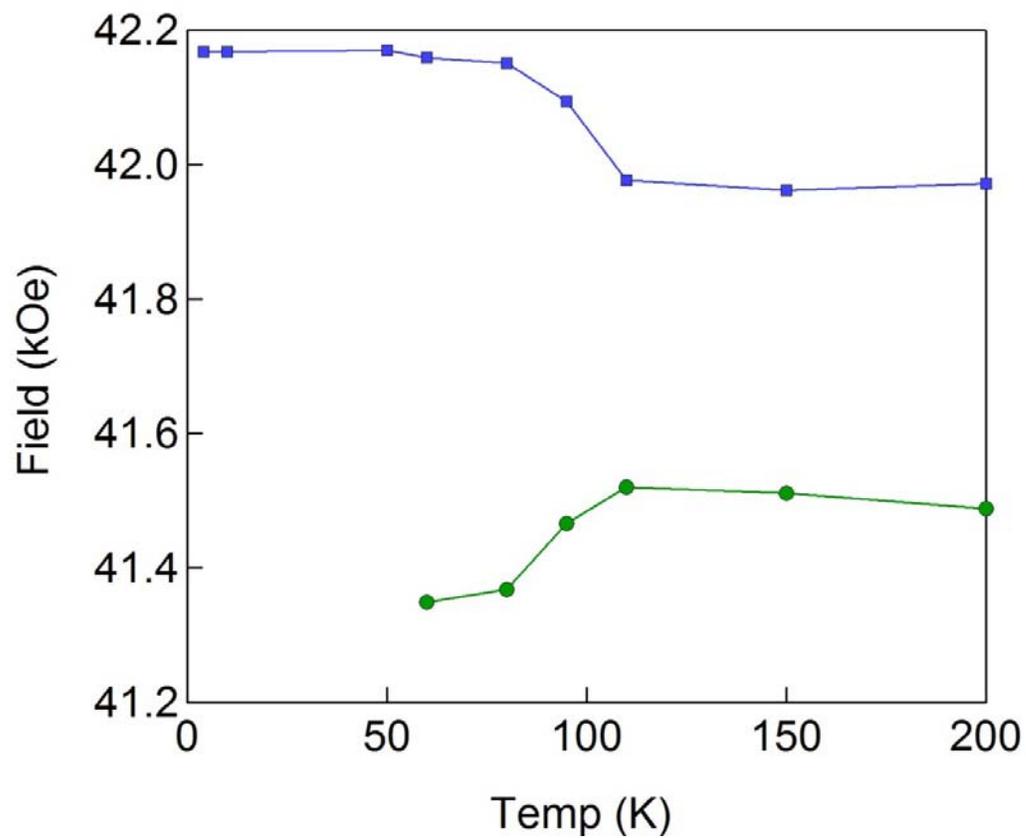

**Fig. S8.** The position of the sub-peaks in the NMR spectra shown in Fig. S7 (indicated by black solid arrows) as a function of temperature.

# References for Supporting Information